\documentclass[10pt,letterpaper]{article}
\usepackage[top=0.85in,left=2.75in,footskip=0.75in]{geometry}

\usepackage{amsmath,amssymb}
\usepackage{wasysym}
\usepackage{multirow}
\usepackage{changepage}

\usepackage[utf8x]{inputenc}

\usepackage{textcomp,marvosym}

\usepackage{cite}

\usepackage{nameref,hyperref}

\usepackage[right]{lineno}

\usepackage{microtype}
\DisableLigatures[f]{encoding = *, family = * }

\usepackage[table]{xcolor}

\usepackage{array}

\newcolumntype{+}{!{\vrule width 2pt}}

\newlength\savedwidth

\raggedright
\usepackage[aboveskip=1pt,labelfont=bf,labelsep=period,justification=raggedright,singlelinecheck=off]{caption}

\makeatletter
\renewcommand{\@biblabel}[1]{\quad#1.}
\makeatother

\usepackage{lastpage,fancyhdr,graphicx}
\usepackage{epstopdf}
\fancyheadoffset[L]{2.25in}
\fancyfootoffset[L]{2.25in}
\begin{document}
\vspace*{0.2in}

\begin{flushleft}
{\Large
\textbf\newline{A spatially uniform illumination source for widefield multi-spectral optical microscopy} % Please use "sentence case" for title and headings (capitalize only the first word in a title (or heading), the first word in a subtitle (or subheading), and any proper nouns).
}
\newline
\\
İris Çelebi\textsuperscript{1\Yinyang},
Mete Aslan\textsuperscript{1\Yinyang},
M. Selim Ünlü\textsuperscript{1,2,3,*}
%with the Lorem Ipsum Consortium\textsuperscript{\textpilcrow}
\\
\bigskip
\textbf{1} Department of Electrical and Computer Engineering Boston University, Boston, MA, USA
\\
\textbf{2} Department of Biomedical Engineering, Boston University, Boston, MA, USA
\\
\textbf{3} iRiS Kinetics Inc, Boston, MA, USA
\\
\bigskip

% Insert additional author notes using the symbols described below. Insert symbol callouts after author names as necessary.
% 
% Remove or comment out the author notes below if they aren't used.
%
% Primary Equal Contribution Note
\Yinyang These authors contributed equally to this work.

% Additional Equal Contribution Note
% Also use this double-dagger symbol for special authorship notes, such as senior authorship.
%\ddag These authors also contributed equally to this work.

%\textcurrency Current Address: Dept/Program/Center, Institution Name, City, State, Country % change symbol to "\textcurrency a" if more than one current address note
% \textcurrency b Insert second current address 
% \textcurrency c Insert third current address


% Group/Consortium Author Note
%\textpilcrow Membership list can be found in the Acknowledgments section.

% Use the asterisk to denote corresponding authorship and provide email address in note below.
* selim@bu.edu

\end{flushleft}
\section*{Abstract}
Illumination uniformity is a critical parameter for excitation and data extraction quality in widefield biological imaging applications. However, typical imaging systems suffer from spatial and spectral non-uniformity due to non-ideal optical elements, thus require complex solutions for illumination corrections. We present Effective Uniform Color-Light Integration Device (EUCLID), a simple and cost-effective illumination source for uniformity corrections. EUCLID employs a diffuse-reflective, adjustable hollow cavity that allows for uniform mixing of light from discrete light sources and modifies the source field distribution to compensate for spatial non-uniformity introduced by optical components in the imaging system. In this study, we characterize the light coupling efficiency of the proposed design and compare the uniformity performance with the conventional method. EUCLID demonstrates a remarkable illumination improvement for multi-spectral imaging in both Nelsonian and Koehler alignment with a maximum spatial deviation of $\approx$ 1\% across a wide field-of-view.

% Please keep the Author Summary between 150 and 200 words
% Use first person. PLOS ONE authors please skip this step. 
% Author Summary not valid for PLOS ONE submissions. 
\bigskip
%\section*{Author summary}
%Lorem ipsum dolor sit amet, consectetur adipiscing elit. Curabitur eget porta erat. Morbi consectetur est vel gravida pretium. Suspendisse ut dui eu ante cursus gravida non sed sem. Nullam sapien tellus, commodo id velit id, eleifend volutpat quam. Phasellus mauris velit, dapibus finibus elementum vel, pulvinar non tellus. Nunc pellentesque pretium diam, quis maximus dolor faucibus id. Nunc convallis sodales ante, ut ullamcorper est egestas vitae. Nam sit amet enim ultrices, ultrices elit pulvinar, volutpat risus.

%\linenumbers

% Use "Eq" instead of "Equation" for equation citations.
\section*{Introduction}
The illumination quality is of paramount importance for optical microscopy.
However, uneven illumination is an issue for imaging techniques, particularly for widefield microscopy applications \cite{ji2022non,herbert2012enhanced,model2001standard}.
In typical imaging systems, both collection and illumination paths are comprised of multiple optical elements that introduce a collection of optical transfer functions that affect the field uniformity.
The most common visible effect of the non-ideal, limited numerical aperture (NA) optics is vignetting \cite{ji2022non,khaw2018flat}. 
For applications such as quantitative fluorescence microscopy uniform excitation of the sample is critical for correct characterization of measured fluorophore response~\cite{mau2021fast}. 
Another application example is optogenetics where spatiotemporal control of uniform and high-power excitation across the sample is crucial~\cite{photonics8110499,repina2020engineered}. 
The optical system we have developed, Interferometric Reflectance Imaging Sensor (IRIS)~\cite{celebi2020instrument}, is another biosensing platform for accurate characterisation of binding kinetics.
For the IRIS system, illumination uniformity also drastically affects the detection accuracy and sensitivity for picometer level increments of biomass accumulation.
Therefore, design of the illumination source that can compensate for non-uniformities in the optical system is essential for numerous imaging and optical sensing applications. 

Critical (or Nelsonian) alignment is one of the simplest illumination configurations where the light source is imaged on the object (sample) plane.
Although this approach provides efficient light coupling in the system, which can be crucial for high signal-to-noise ratio, the underlying limitation is further compromising the light homogeneity. 
In modern light microscopy, Koehler illumination is preferred over critical illumination, where the image of the light source is defocused on the object plane and its conjugate planes. 
Therefore this method provides superior uniformity at the cost of total light power and requires additional optical elements.
However, the common issue of vignetting persists in both configurations.
To compensate for this effect, custom components have been introduced in literature.
Mau et.al. \cite{mau2021fast} proposed a laser scanning technique to have uniform excitation in single molecule localization microscopes, Coumans et.al. \cite{coumans2012flat} introduced microlens arrays to flatten the illumination profile and Model et.al \cite{model2001standard} utilized concentrated fluorophore solutions to correct the spatial heterogeneity of the field.
Computational methods have been developed for image corrections \cite{priya2014retrospective} to avoid introducing optical complexity to the system, however they fail to fully compensate for the excitation field distortion \cite{herbert2012enhanced} and real-time signal readings.  

A particularly challenging microscopy method for illumination design is multi-spectral imaging (MSI), where field uniformity across the different spectral channels plays a key role for accurate interpretation of the signal \cite{sawyer2017evaluation}.
Over the last 25 years, the advances in visible light emitting devices (LEDs) have presented unprecedented capabilities in multi-spectral light sources that provide compact, high-power and user-controlled color illumination.
However, achieving homogeneous excitation for separate wavelength channels, simply due to spatial separation of the source elements, requires complex solutions for the illumination optics. 
The conventional multi-spectral illumination for MSI employs dichroic beamsplitters and/or dichroic mirrors \cite{levenson2006multispectral,hu2020quadrature}. 
These elements require diligent selection of illumination wavelengths and precise optical alignment, as well as they increase the total cost of the entire illumination system.

In this study, we present an LED based light source with an adjustable field profile, termed Effective Uniform Color-Light Integration Device (EUCLID). 
EUCLID, similar to a standard integrating sphere, employs a hollow  cavity with: a diffuse scattering surface, entrance ports for the light source and an exit port. 
The significant novelty of EUCLID is the introduction of conical geometry allowing for design optimization. 
The hollow cavity is engineered to improve the light coupling efficiency and field uniformity for different illumination configurations.
We perform OpticStudio (Zemax) simulations to study the relationship between apex angle of the conical cavity of EUCLID and light output characteristics.
We also examine and compare the field uniformity provided by different light sources under critical and Koehler illumination configurations.
%We also examine and compare the field uniformity provided by the widely used traditional Koehler technique and light sources with the integrating sphere method and the both EUCLID configurations.
The primary finding of this study is the overall improvement in a critical illumination method with both single narrow band and multi-color imaging in comparison to the preferred standard in microscopy. 
We demonstrate the direct impact of EUCLID in imaging quality, providing a remarkable illumination uniformity with $< 1\%$ intensity deviation across different input source channels and $\approx 1\%$ spatial light intensity variation within the full FOV (5mm x 7mm and 2mm x 2.8mm for 2x and 5x objective lenses, respectively) of our system.
We finally test the color integration performance of EUCLID when it is coupled to a multi-mode fiber to demonstrate its applicability for other illumination configurations.

\section*{Materials and methods}
\subsection*{Design of EUCLID}

Interferometric reflectance imaging sensor (IRIS) \cite{celebi2020instrument},  introduces the fundamentals of utilizing diffuse reflections for mixing spatially separated LED sources and achieving a more uniform field profile.
The system employs an integrating sphere in an unconventional manner.
Integrating spheres evenly spread the input light by multiple reflections over the hollow cavity, therefore they are conventionally used for a variety of photometric or radiometric measurements. 
However, the IRIS systems have leveraged the beam produced by multiple diffusive reflection to obtain a light source with constant radiance (W/m$^2$/str) profile for all different color LED sources. 
This technique has also been evaluated for hyperspectral imaging by utilizing a diffuse scattering dome\cite{sawyer2017evaluation}, and has shown as an effective method of achieving spatial homogeneity of discrete light sources.  
Therefore, we obtain effective mixing of multi-spectral LED sources in the conventional IRIS system by utilizing an integrating hollow cavity. 

\begin{figure}[!h]
\includegraphics[scale=1.5]{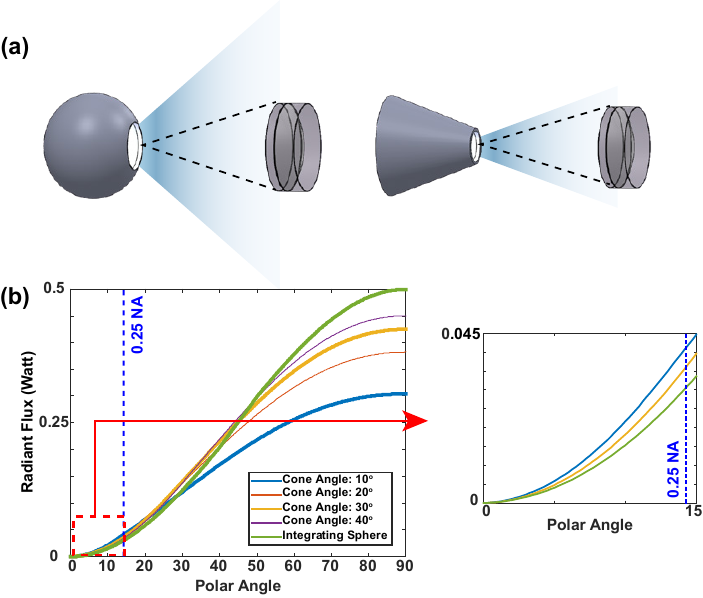}
\caption{{\bf Demonstration of Light Confinement offered by the Conic Geometry.}
The sketch of simulated geometries (a). Output radiant flux vs. polar angle graph obtained from OpticStudio simulations. Supplied power to the LIDs is set to 1 Watt.}
\label{fig:simulation}
\end{figure}

In IRIS systems, the sample is illuminated with common path reflectance mode, with either Koehler \cite{daaboul2010high} or critical \cite{celebi2020instrument} illumination techniques.
Although the color mixing is achieved, the overall efficiency of light coupling is reduced given the source profile (constant radiance) is not optimized for the finite-NA illumination optics.
We have selected a conical hollow cavity geometry for EUCLID, to concentrate the light within the illumination path NA.
The conical structure provides an intuitive geometry to confine the output rays into a prescribed cone.
The geometry also enables a simple path for further modification of the cavity, since the light output emanates predominantly from the scattering on the base of the cone.
The effect of conical geometry was studied by performing Non-Sequential ray tracing analysis on OpticStudio as it is described in \textit{OpticStudio Simulations} section.
Figure \ref{fig:simulation}a shows the spherical and conical geometries and Figure \ref{fig:simulation}b demonstrates the improvement of power confinement for a 0.25NA condenser lens.
After the light coupling analysis and validation of simulations for the conical cavity, we modify the geometry further, to shape the output field profile to achieve improved uniformity. 
In an aberration-free imaging system, irradiance of point $A$ on the image plane, $E_A$ is defined by
\begin{equation}\label{irradianceEQN}
    E_A = \int L(\theta,\phi)\,cos(\theta)\,d\Omega
\end{equation}
where $L(\theta,\phi)$ is the radiance and the limit of this integral is determined by the physical limits of the exit pupil.
If the radiance in Eq.\ref{irradianceEQN} is constant, i.e., Lambertian source is used, then irradiance at point $A$ is proportional to the projected solid angle subtended by the exit pupil according to point $A$~\cite{10.1117/12.938414}.
Thus, for an on-axis point, the exit pupil spans more area in the directional cosine space than for an off-axis point, which results to a bright spot at the center of the image plane.
The irradiance uniformity degrades even further if other aberration causes, such as cosine-fourth law and pupil aberration, are considered~\cite{937554}.
To mitigate this issue, complex illumination source and lens designs are introduced and varying aperture configurations are studied in the literature~\cite{coumans2012flat,ji2022non}.

\begin{figure}[!h]
\includegraphics[scale=2]{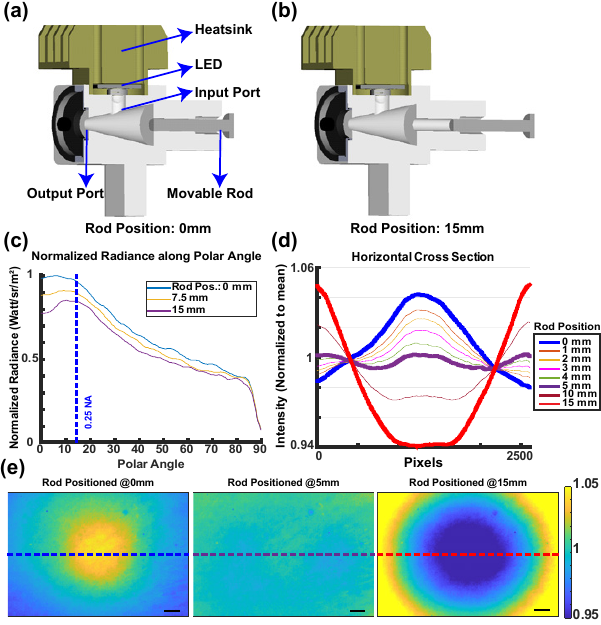}
\caption{{\bf Design and Adjustable Field Performance of the EUCLID.}
3D model of EUCLID, adjustable rod positioned at 0mm (a) and 15mm (b). The effect of adjustable rod position on the output radiance profile (c). The cross section (d) and heatmaps (e) of intensity images of the sample with varying rod positions. Rod positioned at 0mm corresponds to nominally flat base surface of the cone. The scale bar is 500$\mu$m.}
\label{fig:adjustable}
\end{figure}

The simple design of EUCLID however, allows intuitive design optimization without further complicating the optics.
Thanks to its conical geometry and small output port dimension, the radial distribution of the output depends only the lights scattered from the back surface (base of the cone). 
Thus, with engineering the back surface structure, the output radiance can be altered such that the finite-size aperture and aberration effects can be alleviated to achieve a uniform field profile. 
We opted for a design given in Fig.\ref{fig:adjustable} where the output radiance can be controlled and changed with a movable rod for different imaging systems with different exit pupil sizes. Note that the rod material is identical to the hollow cavity material.

The geometry of the cavity shape indicate that the majority of output light is scattered from the base of the cone.
Therefore, the diameter of the movable rod, namely the diameter of the guiding hole on the back surface, given in Fig.\ref{fig:adjustable}, affects the output radiance of EUCLID. 
It is possible to define the lower bound of output rays angles, $\theta_p$, when the rod position is at infinity, effectively creating a hole on the back surface.
Using geometrical arguments (see S4 Fig.), this pass angle, $\theta_p$, can be defined as
\begin{equation}
    \theta_p = tan^{-1}\Big(\frac{\frac{D}{2}-\frac{\diameter_{\text{output}}}{2}}{h}\Big)
\end{equation}
where $D$ and $\diameter_{\text{output}}$ are the diameters of guiding hole and output port, respectively and $h$ is the height of the conical cavity.
In the presence of the movable rod, the output radiance can be tuned, and this characteristic of EUCLID compensates the drop in the high-frequency components of the illumination transfer function, typically due to finite-sized circular lenses employed in imaging systems and optical aberrations. 
In Koehler configuration, this pass angle and the aperture stop of the system determines the area in which the profile can be altered. 
In critical illumination, it is preferable to match the output dimension and the rod diameter for small aperture systems, since it gives the flexibility to adjust the distribution of the field profile with changing the rod position while minimizing the power loss due to the hole.

%%%%%%%%%%%IRIS- can we mention the cost somewhere else? Here its all about design not cost effectiveness
We have fabricated a spherical, 23$^{\circ}$ and 15$^{\circ}$ apex angle conical LIDs from high reflectivity PTFE blocks (reflectivity $\sim$ 96$\%$).
The guiding hole and rod diameter is selected to be 7.9mm for EUCLID to affect the field profile in both illumination configurations. 

% For figure citations, please use "Fig" instead of "Figure".

% Place figure captions after the first paragraph in which they are cited.

% Results and Discussion can be combined.
\subsection*{OpticStudio Simulations}
%%%%%%%%%%IRIS - should we show an image right here? Show mete what you mean
The effect of conical geometry was studied by performing Non-Sequential ray tracing analysis on OpticStudio for various light integrating devices (LIDs).
For this purpose, we created 7 different conical light integrating devices (LIDs) and 1 integrating sphere whose port fractions, i.e. the ratio of output port and internal area, were designed to be identical.
The apex angle of the conical LIDs was swept from $10^\circ$ to $30^\circ$ with a $5^\circ$ step size and the output port diameter was selected as 5 mm for all devices.
We used built-in objects to construct the LIDs and IS, and internal coatings were set to have a Lambertian scattering profile with 99\% reflectivity.
We placed two rectangular detectors on the back surface and output of the conical LIDs to validate light distributed evenly inside the conical structure.
The total radiant flux within the far-field polar angles was calculated from the output light distribution, measured on the polar detector with a radius much larger than the output port dimensions ($r_{\text{polar detector}}=60$ mm). We have also measured the output radiance of each LID from the rectangular detector placed on the output ports.
We also couple EUCLID to a multi-mode fiber and achieve uniform output field profiles for different color channels. 
This indicates that EUCLID can be employed in applications where spatiotemporal control of uniform excitation is crucial, such as optogenetics.
With temporal control of the different color LED input stimulation, EUCLID can also provide multi-color pulsed illumination and be employed in optogenetics applications since it offers high power illumination with even distribution on wide FOVs.

\section*{Results and Discussion}
\subsection*{Power Efficiency due to Conical Geometry}
\begin{figure}[!h]
\includegraphics[scale=0.8]{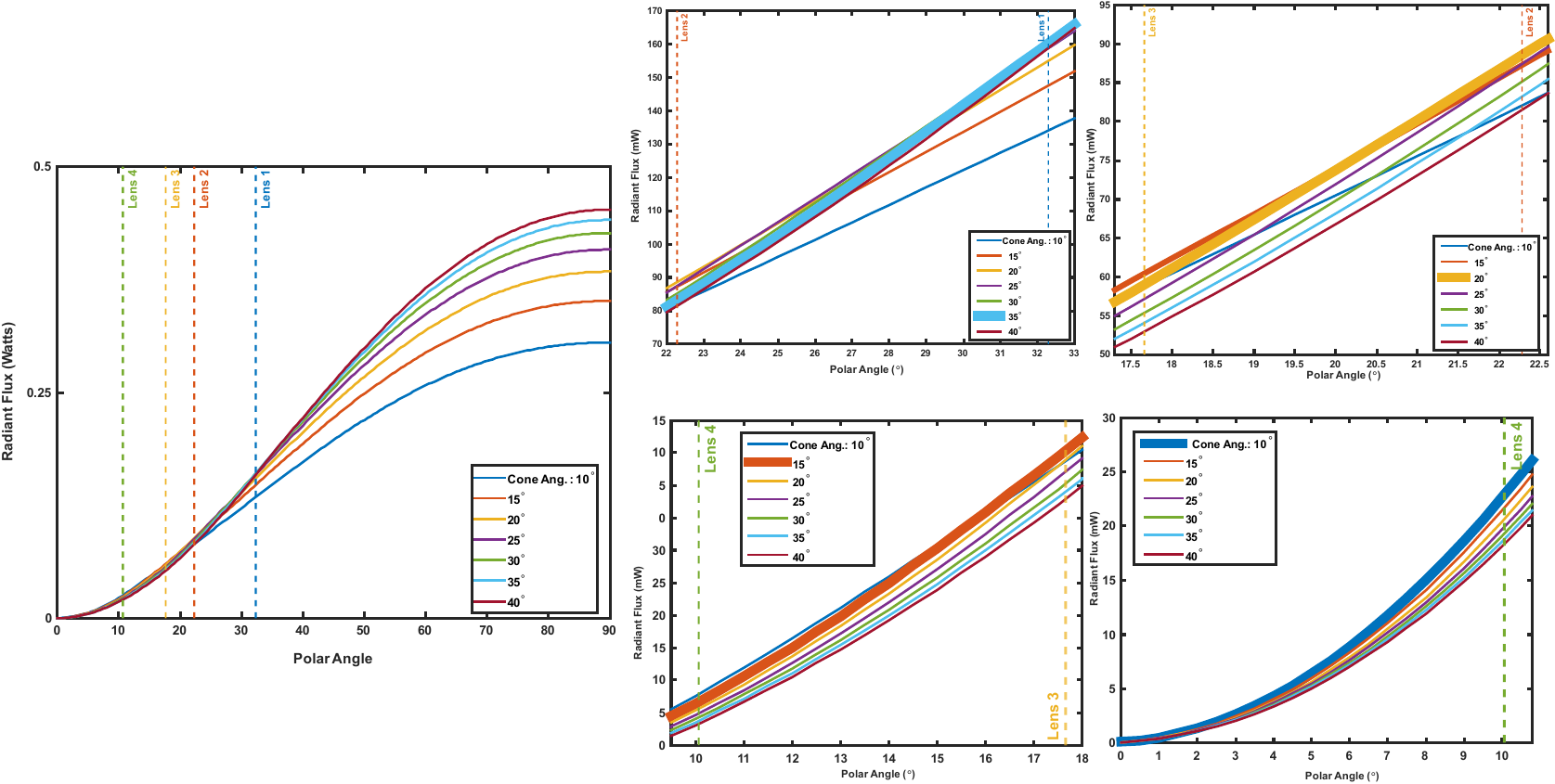}
\caption{{\bf Power Coupling Simulation Results.}
Total output radiant flux vs. output polar angles graph calculated by Non-sequential ray tracing simulations. Left: Full output characteristic of various conic LIDs. Right: Zoomed sections of the left figure. Best performing conic LID is indicated with the bold lines. The acceptance angles for the lenses are 32.29$^\circ$, 22.28$^\circ$, 17.66$^\circ$ and 10.70$^\circ$ for lens 1, 2, 3 and 4, respectively.}
\label{fig:ZemaxPowCoupSim}
\end{figure}

We have first analyzed the light confinement characteristic of conic geometry by OpticsStudio simulations.
As the simulation parameters are explained in the \textit{OpticStudio Simulations} section, we determined optimal apex angles for different commercial condenser lenses employed in 4-f Koehler system or critical illumination configuration.
As it is given in Fig.\ref{fig:ZemaxPowCoupSim}, the total coupled optical power is maximized when the apex angle of EUCLID matches with the acceptance angle of the first condenser lens.

To test and compare the experimental light coupling performances of EUCLID and IS, we have built a conventional IRIS system \cite{celebi2020instrument} with a 50mm condenser lens in the illumination path (See S1 Fig.).
In the imaging system, we employed a 2X 0.06NA (CFI Plan Achro) objective lens to focus the illumination light on the a reflective silicon substrate and image the sample plane with a CMOS camera (BFS-U3-70S7M-C).
The mean CMOS readings were recorded for analyzing light coupling performances of LIDs under identical experimental settings. 
Table \ref{tab:LightCoupling} shows the measurement readings and normalized values.
Although the total output flux from the conical LIDs are decreased, the confined flux within the system is increased by 28\% and 58\% for 23$^{\circ}$ conical and 15$^{\circ}$ conical LIDs, respectively.
We fabricated our proof-of-concept LIDs from a PTFE block (reflectivity 96$\%$) which cost under \$50. 
However, the output power of LIDs can be increased using commercial materials or coatings with higher reflectivity around 99$\%$ (i.e. Spectralon - LabSphere, Flourilon - Avian Technologies).
\begin{table}[!ht]
%\begin{adjustwidth}{-2.25in}{0in} % Comment out/remove adjustwidth environment if table fits in text column.
\centering
\caption{\bf Light Coupling Performances of LIDs}
\begin{tabular}{|c|c|c|c|c|c|}
\hline
Device  & Output & Power  & Mean CMOS  & Exp. & Reading \\
 & Power (mW)& Ratio (/IS) & Reading & Time (ms) & Ratio (/IS)\\
\hline
IS & 102.4 & 1 & 12770 & 0.5 &  1\\
23$^{\circ}$ Conic & 91.8 & 0.9 & 16361 & 0.5 & 1.28\\
15$^{\circ}$ Conic & 76.5 & 0.75 & 20180 & 0.5 & 1.58\\
\hline
\end{tabular}
  \label{tab:LightCoupling}
%\end{adjustwidth}
\end{table}

\subsection*{Adjustable Field Profile}
We have simulated the effect of the rod position on output radiance.
As the results in Figure \ref{fig:adjustable}c (also S2 Fig.), indicate, output radiance can be fully controlled by changing the rod position which is desired to compensate cosine-fourth and vignetting effects for different imaging systems.
Finally, to experimentally validate the performance of EUCLID, we have illuminated and imaged a flat SiO$_2$-Si substrate with the setup (see S1 Fig.) and achieved an ultimate profile uniformity with 1.01\% min-max deviation along the horizontal cross section (Figure \ref{fig:adjustable}d).

\begin{figure}[!h]
\includegraphics[scale=1.1]{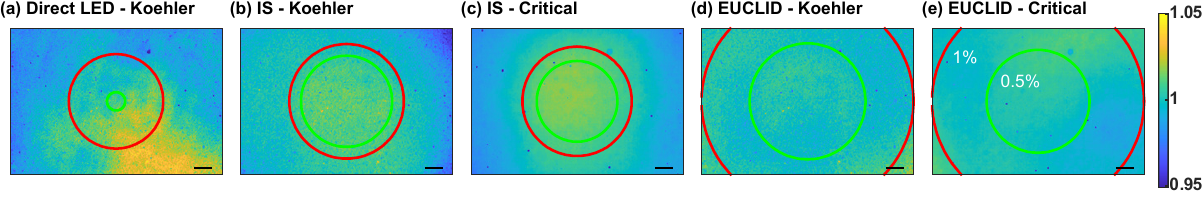}
\caption{{\bf Uniformity regions of different illumination sources.}
The red and green circles indicate the area where the irradiance profile deviation is $< 1\%$ and  $< 0.5\%$, respectively. The uniform irradiance circles are shown for: direct LEDs (a), spherical LID (b), EUCLID (d) in Koehler alignment; spherical LID (c), EUCLID (e) in critical alignment.
The scale bar is 200$\mu$m.}
\label{fig:Uniformity}
\end{figure}

%% Uniformity Performances of Light Sources
To test the uniformity performance of EUCLID, we have compared different illumination configurations for both critical and Koehler alignment.
In this experimental setup (see S1 Fig.) a 5x 0.15NA (TU Plan Flour) objective and a narrow band LED were used. 
All of the LIDs have identical output port diameter, resulting in identical illumination area.
In order to quantify field uniformity for the resulting raw images, we have defined a parameter termed `uniformity regions', where the normalized intensity deviation is under a certain threshold (for instance 1\% or 0.5\%). 
The radii of these regions provide a quantitative metric to compare performance of different illumination configurations and how these regions are determined is explained in the S1 Appendix.  
The uniformity regions for all sources are illustrated in Figure \ref{fig:Uniformity} and the calculated radii of 1\% and 0.5\% uniformity regions are given in Table \ref{tab:UniformityRegions}.

\begin{table}[!ht]
\centering
\caption{\bf Radii of Uniformity Regions}
\begin{tabular}{|m{5em} |m{4.8em}|m{3.5em}|m{4.8em}|m{3.5em}|}
\hline
\multirow{2}{*}{\bf Source} & \multicolumn{2}{c|}{\bf Koehler} & \multicolumn{2}{c|}{\bf Critical} \\
\cline{2-5}
 & \bf 1\% &\bf  0.5\% & \bf 1\% & \bf 0.5\% \\
 \hline
{\bf Direct LED} & {580 pxl} & {110 pxl} & {-} & {-} \\
{\bf IS} & {705 pxl} & {560 pxl} &{670 pxl} & {495 pxl} \\
{\bf EUCLID} & Full FOV (1581 pxl)& {710 pxl} & Full FOV (1581 pxl)& {630 pxl}\\
\hline
\end{tabular}
  \label{tab:UniformityRegions}
\end{table}

EUCLID, when the rod is positioned at 2mm and 6mm away from the back surface for Koehler and critical setup respectively, outperforms the integrating sphere and the standard approach of direct LEDs with a Koehler setup.
Our design succeeded to obtain a 1.05\% max-min deviation across the entire FOV.

\subsection*{Multi-spectral Illumination and Fiber Coupling}
\begin{figure}[!h]
\includegraphics[scale=0.75]{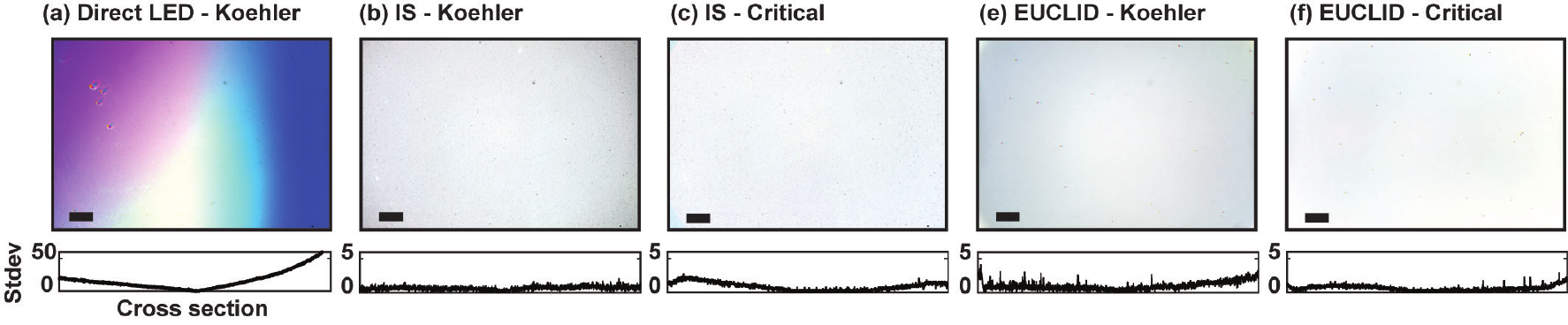}
\caption{{\bf Pseudo RGB images (top) and respective spectral deviation, across the horizontal cross section (bottom).}
Direct LEDs (a), Spherical LID (b), EUCLID (d) in Koehler alignment; Spherical LID (c), EUCLID (e) in critical alignment. 
The scale bar is 200$\mu$m.}
\label{fig:colormix1}
\end{figure}

Uniform illumination profile for different wavelength channels is essential for MSI techniques. 
The integrating hollow cavity allows mixing and a uniform output from spatially separated LED dies. 
We have demonstrated the imaging improvement using an integrating cavity, with a 3-channel multi-spectral image cube we acquired from separate color channels.
Using a monochrome camera with fixed optics, we have sequentially turned on different LED channels (633nm dominant, 517nm dominant, 453nm dominant, OSRAM LZ series) and acquired separate images. 
We then created the pseudo RGB images after normalizing each channel to account for variations in die brightness.
The pseudo RGB images were created for five different illumination path conditions: direct LEDs, spherical LID, EUCLID in Koehler alignment; spherical LID, EUCLID in critical alignment.
We have analyzed the MSI uniformity performances by examining the brightness deviations with respect to other channels across the FOV.
The standard deviation along the color channel dimension was recorded for each pixel.
Figure \ref{fig:colormix1} shows the horizontal cross sectional view of color deviation for each condition. 
As expected, without the use of an integrating device, we observed a poor color mixing performance and effectively a reduced uniformly illuminated FOV.
EUCLID in critical illumination configuration showed the best performance for color uniformity with an average deviation of less than 0.6\%.

\begin{figure}[!h]
\includegraphics[scale=1.5]{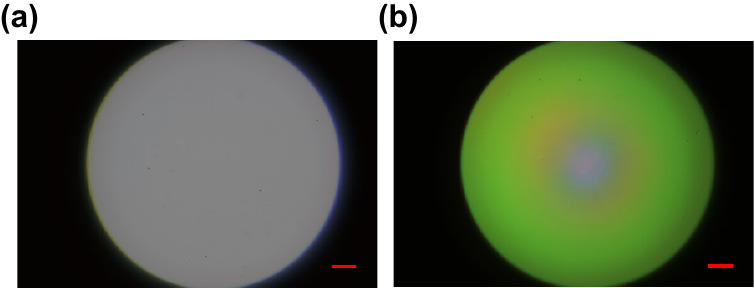}
\caption{{\bf Pseudo RGB images of fiber facet.}
Output of EUCLID (a) and direct LEDs (b) are coupled to the input end of the fiber. The scale bar is 200$\mu$m}
\label{fig:fiberCopuling}
\end{figure}

%As a final evaluation of color mixing performance of EUCLID, we studied light coupling 
Multi-mode fibers can be used for compact and convenient multi-spectral light delivery in microscopy \cite{meng2018spectrally}.
We studied the spectral mixing performance of EUCLID for fiber-optic applications, as an additional evaluation.
We coupled the light sources to the fiber facet (Thorlabs M93L02, $\diameter_{\text{core}} = 1.5$mm, 0.39NA), and imaged the output of the fiber to analyze field profiles with respect to separate source channels (See S4 Fig.).
We used a lens pair, 200mm and 30mm, to couple the light sources into the core, by matching the fiber NA.
The system was aligned for the 453nm dominant channel for both illumination systems and the optical alignment was kept fixed.
We then acquired images for the remaining spectral channels and created the psuedo RGB images shown in Figure \ref{fig:fiberCopuling}.
EUCLID maintained a uniform spatial and spectral profile. The direct LED coupling suffered from spatial separation of LED dies, resulting in an uneven field with peripheral and central concentration of separate channels.

%PLOS does not support heading levels beyond the 3rd (no 4th level headings).

%\section*{Discussion}
%Nulla mi mi, venenatis sed ipsum varius, Table~\ref{table1} volutpat euismod diam. Proin rutrum vel massa non gravida. Quisque tempor sem et dignissim rutrum. Lorem ipsum dolor sit amet, consectetur adipiscing elit. Morbi at justo vitae nulla elementum commodo eu id massa. In vitae diam ac augue semper tincidunt eu ut eros. Fusce fringilla erat porttitor lectus cursus, vel sagittis arcu lobortis. Aliquam in enim semper, aliquam massa id, cursus neque. Praesent faucibus semper libero~\cite{bib3}.

\section*{Conclusion}

In summary, we have demonstrated the light coupling and multi-spectral illumination performance of EUCLID.
We have validated the feasibility of a diffusive scattering hollow cavity as the light source in a widefield imaging system, and an effective spatial uniformity correction. 
The intuitive conical geometry and an adjustable cavity serves the purpose of introducing a novel design parameter for light illumination devices.
EUCLID and variations of light integrating devices with further modifications in the geometrical design, can accommodate and correct for imaging systems with different effective transfer functions, elements and alignment. 
With temporal control of the different color LED input stimulation, EUCLID can also provide multi-color pulsed illumination.

\section*{Supporting information}

% Include only the SI item label in the paragraph heading. Use the \nameref{label} command to cite SI items in the text.
\paragraph*{S1 Fig.}
\label{S1_Fig}
\includegraphics[scale=0.7]{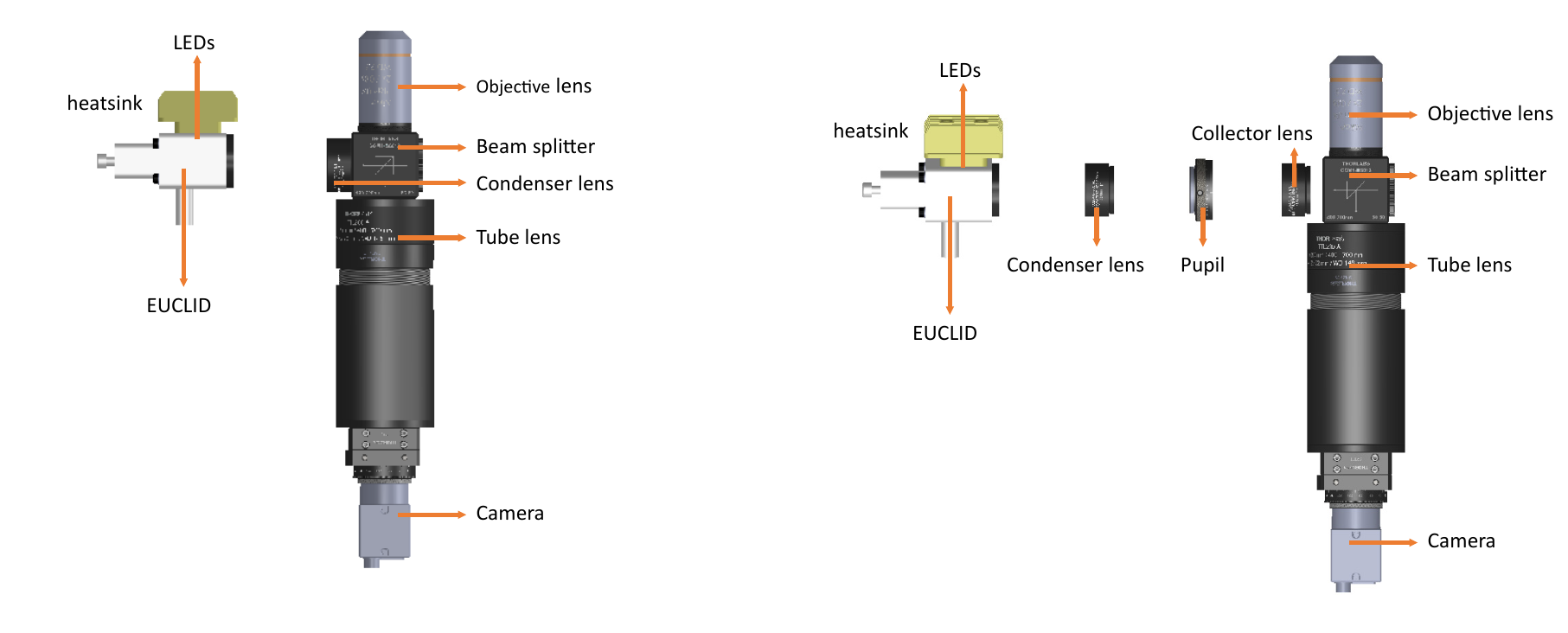}
{\bf Optical Setup.} Detailed schematics of the imaging setups. Left: with critical illumination, Right: Koehler illumination.
The light integration devices (LIDs) or direct LED dies were aligned to the imaging optics in identical conditions to acquire field profiles. 

\paragraph*{S2 Fig.}
\label{S2_Fig}
\includegraphics[scale=0.7]{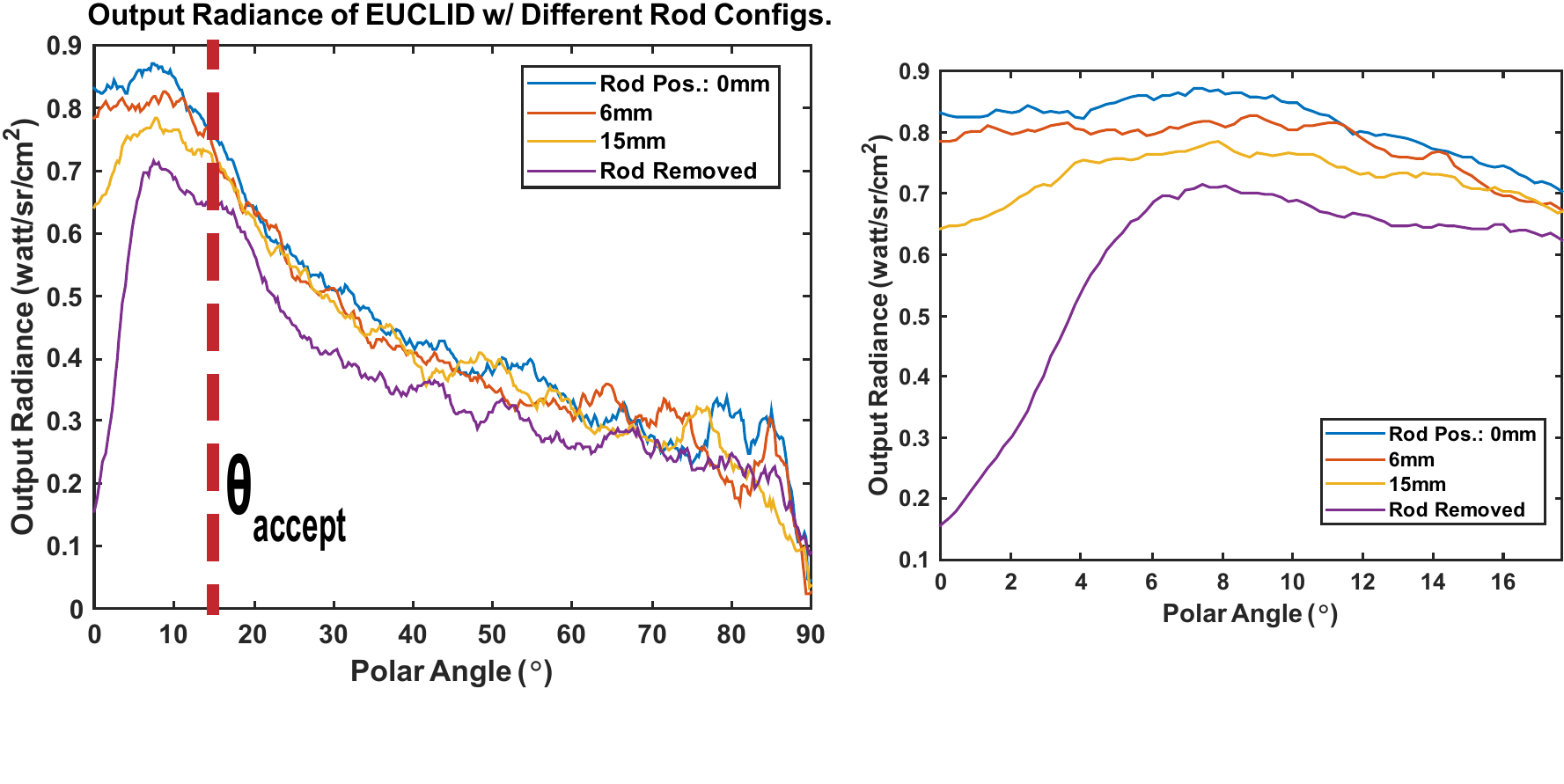}
{\bf Output Radiance Simulations.} Output Radiance cross sections of EUCLID with 5 mm output port and rod diameter when rod is positioned different locations. Left: Output radiance for all polar angles. Acceptance angle is defined by lens 3 in Fig.\ref{fig:ZemaxPowCoupSim}. Right: Zoomed section of left graph for angles that lies within the acceptance angle.

\paragraph*{S3 Fig.}
\label{S3_Fig}
\includegraphics{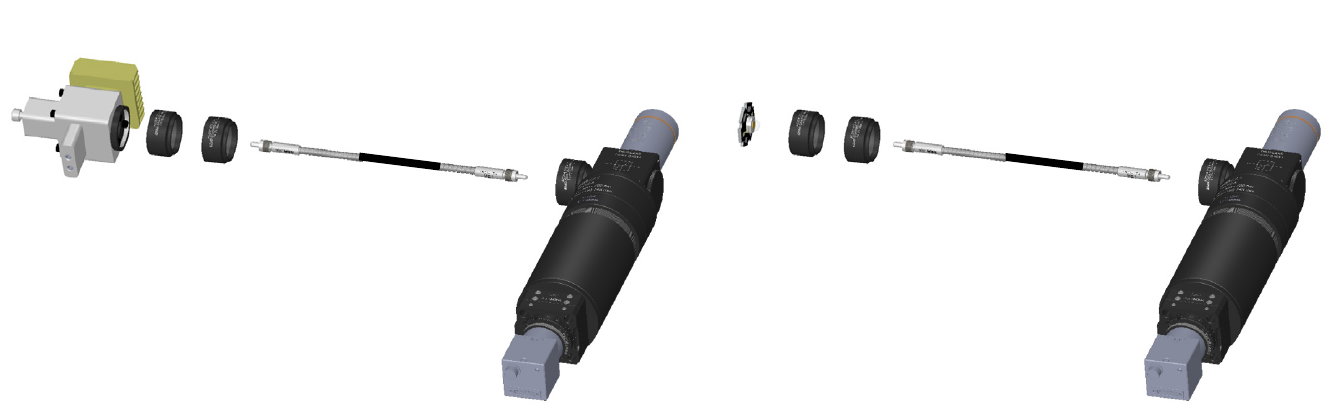}
{\bf Fiber Performance Setup.} Schematic of the imaging setups for fiber alignment, with direct LEDs (right) and with EUCLID (left).
The light output of direct LED dies and the EUCLID were coupled to the fiber tip by a lens pair for demagnification. The output of the fiber tip is then imaged by using the same collection optics as previous setups. 

\paragraph*{S4 Fig.}
\label{S4_Fig}
\includegraphics[scale=0.9]{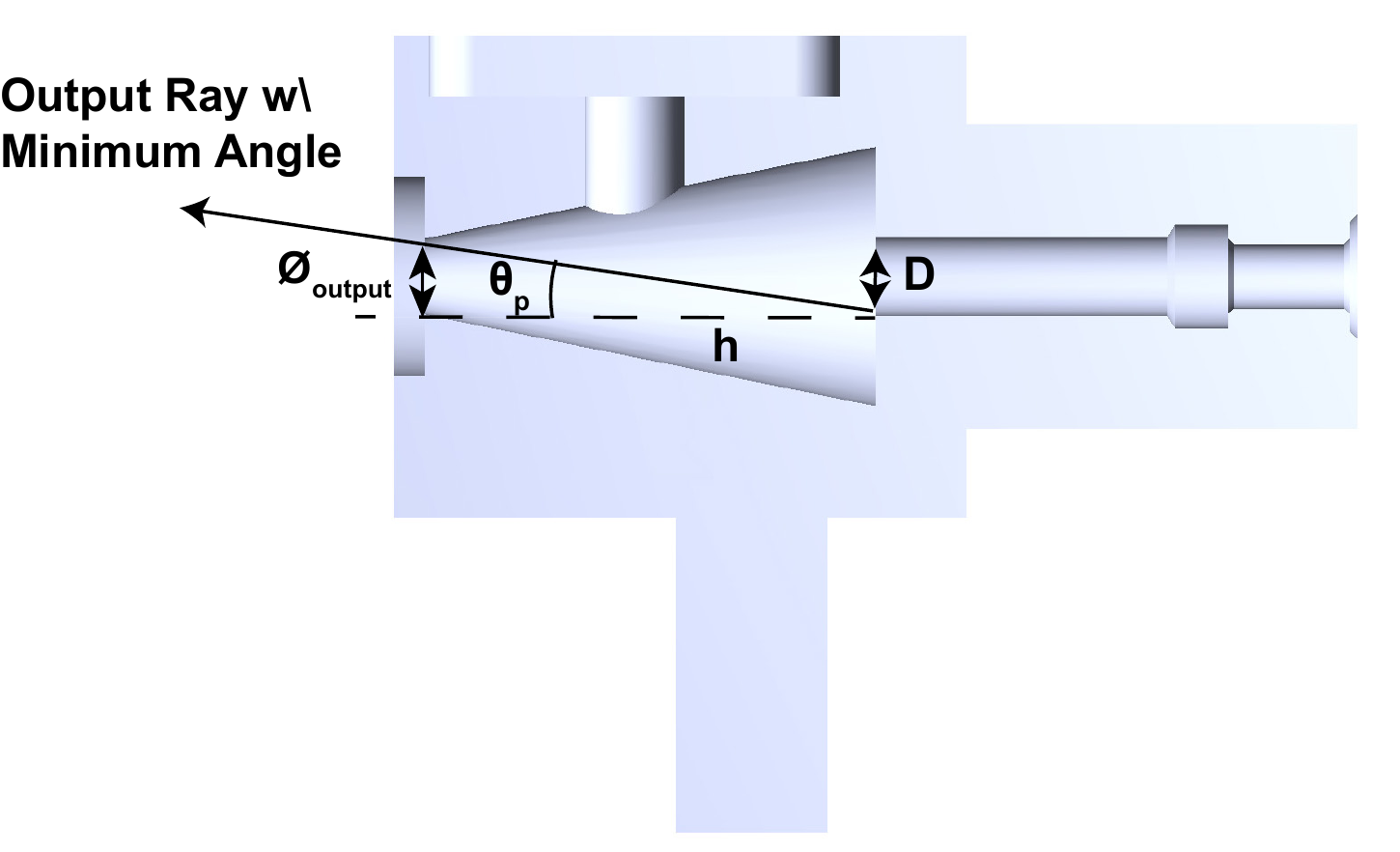}
{\bf Ray approach for EUCLID.} Toy picture of the EUCLID geometry where the output ray with the minimum exit angle is indicated. $h$ is the height, $\diameter_{\text{output}}$ is the output port and $D$ is the guiding hole diameter of the EUCLID.

\paragraph*{S1 Appendix.}
\label{S1_Appendix}
{\bf Custom MATLAB Code to Determine Uniformity Regions.} To quantify the illumination profile uniformity, we have defined uniformity regions as the largest area enclosed, that yield a normalized intensity deviation under 1\% or 0.5\%.
We calculated the radii of the regions with a custom MATLAB algorithm.
The algorithm initializes the ROI with a circular region (radius of 25 pixels) and compares the mean values of the current ROI to the annulus (with a thickness of 10 pixels) just outside of the initial region.
If the average absolute brightness deviation of the annulus, compared to the previous ROI, is within the specified range (1 or 0.5\%) the algorithm continues to evaluate the absolute brightness deviation in the next annulus with the same thickness, just outside of the previous annular ROI.
The algorithm iterates until the specified annular ring indicates that the uniformity level has decreased below the specified range or the radii hits the corner of the FOV.
The uniformity regions for all sources are illustrated in Fig.\ref{fig:Uniformity} and the calculated radii of 1\% and 0.5\% uniformity regions are given in Table \ref{tab:UniformityRegions} in the manuscript. 
We also applied Gaussian filter ($\sigma=5$) to remove the small artefacts on the sample, such as dust particles, etc.

\section*{Acknowledgments}
The authors acknowledge Nevzat Yaraş of iRiS Kinetics for valuable help in the fabrication of LIDs. 

\nolinenumbers

% Either type in your references using
% \begin{thebibliography}{}
% Text
% \end{thebibliography}
%
% or
%
% Compile your BiBTeX database using our plos2015.bst
% style file and paste the contents of your .bbl file
% here. See http://journals.plos.org/plosone/s/latex for 
%\begin{thebibliography}{10}

%\bibitem{bib1}
%Conant GC, Wolfe KH.
%\newblock {{T}urning a hobby into a job: how duplicated genes find new
%  functions}.
%\newblock Nat Rev Genet. 2008 Dec;9(12):938--950.

%\bibitem{bib2}
%Ohno S.
%\newblock Evolution by gene duplication.
%\newblock London: George Alien \& Unwin Ltd. Berlin, Heidelberg and New York:
%  Springer-Verlag.; 1970.

%\bibitem{bib3}
%Magwire MM, Bayer F, Webster CL, Cao C, Jiggins FM.
%\newblock {{S}uccessive increases in the resistance of {D}rosophila to viral
%  infection through a transposon insertion followed by a {D}uplication}.
%\newblock PLoS Genet. 2011 Oct;7(10):e1002337.

%\end{thebibliography}
\bibliography{main.bib}

\end{document}